\documentclass[aps,twocolumn,superscriptaddress]{revtex4-1}
\usepackage{graphicx,color}
\usepackage{amsmath}
\usepackage{amsfonts}
\usepackage{amssymb}
\usepackage{amsthm}
\usepackage{hyperref}
\usepackage{bm}

\newcommand{\ket}[1]{|{#1}\rangle}
\newcommand{\bra}[1]{\langle{#1}|}
\newcommand{\Tr}{\mathop{\mathrm{Tr}}\nolimits}
\newcommand{\ad}{\mathop{\mathrm{ad}}\nolimits}
\newcommand{\Ad}{\mathop{\mathrm{Ad}}\nolimits}

\newcommand{\beq}{\begin{equation}}
\newcommand{\eeq}{\end{equation}}

\definecolor{dgreen}{rgb}{0,0.5,0}

\definecolor{delete}{cmyk}{0.5,0,0,0}

\begin{document}
\author{Christian Arenz} 
\affiliation{Institute of Mathematics, Physics, and Computer Science, Aberystwyth University, Aberystwyth SY23 2BZ, UK}
\affiliation{Frick Laboratory, Princeton University, Princeton NJ 08544, US}
\author{Daniel Burgarth} 
\affiliation{Institute of Mathematics, Physics, and Computer Science, Aberystwyth University, Aberystwyth SY23 2BZ, UK}
\author{Vittorio Giovannetti}
\affiliation{NEST, Scuola Normale Superiore and Istituto Nanoscienze-CNR, I-56126 Pisa, Italy}
\author{Hiromichi Nakazato}
\affiliation{Department of Physics, Waseda University, Tokyo 169-8555, Japan}
\author{Kazuya Yuasa}
\affiliation{Department of Physics, Waseda University, Tokyo 169-8555, Japan}
\title{Lindbladian purification}
\date{\today}

\begin{abstract}
In a recent work [D. K. Burgarth \textit{et~al.},  Nat.\ Commun.\ \textbf{5},  5173 (2014)] it was shown that a series of frequent measurements can project the dynamics of a quantum system onto a subspace in which the dynamics can be more complex. In this subspace even full controllability can be achieved, although the controllability over the system before the projection is very poor since the control Hamiltonians commute with each other.
We can also think of the opposite: any Hamiltonians of a quantum system, which are in general noncommutative with each other, can be made commutative by embedding them in an extended Hilbert space, and thus the dynamics in the extended space becomes trivial and simple.
This idea of making noncommutative Hamiltonians commutative is called ``Hamiltonian purification.''
The original noncommutative Hamiltonians are recovered by projecting the system back onto the original Hilbert space through frequent measurements.
Here we generalize this idea to open-system dynamics by presenting a simple construction to make \emph{Lindbladians}, as well as Hamiltonians, commutative on a larger space with an auxiliary system.
We show that the original dynamics can be recovered through frequently measuring the auxiliary system in a non-selective way. 
Moreover, we provide a universal pair of Lindbladians which describes an ``accessible'' open quantum system for generic system sizes. 
This allows us to conclude that through a series of frequent non-selective measurements a nonaccessible open quantum system generally becomes accessible. 
This sheds further light on the role of measurement backaction on the control of quantum systems.
\end{abstract}
\maketitle

\section{Introduction} 
Noncommutativity is one of the key features of quantum mechanics. 
The order in which operations and/or measurements are performed influences the outcomes of an experiment. 
In particular in the Lie-theoretical approach to quantum control theory \cite{LieTheoryQuantumC} the noncommutativity plays an important role. 
The goal of quantum control is to steer a quantum system to realize a desired transformation on it by shaping classical time-dependent fields 
\cite{ControlRev}.
Here the noncommutativity of the generators associated with the control fields influences the complexity of the resulting dynamics.
For instance, for two commuting Hamiltonians $H_1$ and $H_2$, which can be switched on and off by external control fields, the resulting unitary evolution is just equivalent to the one generated by a linear combination of $H_1$ and $H_2$.
{On the contrary, by properly concatenating  transformations induced by two noncommuuting Hamiltonians one  can  produce effective evolutions associated with generators which are linearly independent of the original ones, enabling the system 
to explore more ``directions.'' 
For a finite-dimensional closed quantum system $Q$, 
the set of effective  evolutions that can be implemented in this way is formed by the  unitaries  of the Lie group $e^{\mathfrak L({\mathbf H})}$
generated by the dynamical Lie algebra  $\mathfrak L({\mathbf H})$ associated to the set of control Hamiltonians ${\mathbf H} := \{ H_1, H_2, \ldots\}$, 
i.e., the real vector space spanned by all possible linear combinations of the elements of ${\mathbf H}$ and their iterated commutators \cite{LieTheoryQuantumC, AccessibilityandControl, BookDalessandro}.
Accordingly a close system $Q$ characterized by a control set ${\mathbf H}$ 
is said to be fully controllable if 
 $e^{\mathfrak L({\mathbf H})}$ includes all possible unitary transformations on $Q$, or equivalently, 
if the dynamical Lie algebra $\mathfrak L({\mathbf H})$
spans the whole  operator algebra of  $Q$, this last property of ${\mathbf H}$ being also referred to as accessibility.}

{When it comes to open quantum systems, the characterization of 
the reachable (realizable) operations, as well as the associated notion of controllability, becomes more complicated since the allowed operations do not possess a group structure and the notion of dynamical generators is  typically lost \cite{OpenQSC3, OpenQSC1}. A partial exception is provided by the subset of Markov processes which
are equipped with a semigroup structure and admit the notion of dynamical  generators, i.e., the super-operators of  Gorini-Kossakowski-Lindblad-Sudarshan (GKLS) form \cite{KurniawanThesis, Kurniawan} (Lindbladians in the following). Still also in this  case determining which dynamics can be activated by controlling a given collection ${\mathbf L}:=\{ \mathcal{L}_1, \mathcal{L}_2, \ldots \}$  of Lindbladians  is 
a difficult unsolved problem.
One would be tempted to tackle it by  studying the Lie algebra  $\mathfrak L({\mathbf L})$ generated by ${\mathbf L}$ and the corresponding Lie group $e^{\mathfrak L({\mathbf L})}$. However, at variance with the closed system scenario,  linear combinations and commutators of elements of ${\mathbf L}$ will in general produce super-operators which
are no longer allowed  dynamical generators  (e.g., they cannot be cast in the GKLS form), or said it differently, $e^{\mathfrak L({\mathbf L})}$ will include  
transformations which are unphysical. Furthermore,  even for the elements of $e^{\mathfrak L({\mathbf L})}$  which are physically allowed,
  it is in general not clear if it would be possible to implement them by simply playing with  the control fields.
In view of these facts  for open quantum systems one distinguishes the physical notion of controllability, i.e., the ability of using  ${\mathbf L}$ and the classical fields which activate them to perform all physically allowed quantum transformations, from the weaker notion of accessibility, which in this case corresponds to have $\mathfrak L({\mathbf L})$ equal to whole Lie algebra generated by arbitrary Lindbladians. In the following, we will refer to this as the `GKLS algebra' noting however that it contains many elements which are not of GKLS form. 
Differently from the closed quantum system case, it is indeed possible that a control set ${\mathbf L}$  is accessible but not controllable.
Still studying the accessibility of a collection of Lindbladians  is a well posed  mathematical problem, which 
can also shed light on the controllability issue, with accessibility being a necessary condition for controllability. Furthermore accessibility implies that the reachable set has 
non-zero volume and therefore has physical relevance: the short time dynamics explores a high dimensional space and is therefore of high complexity.

It turns out that almost all control sets ${\mathbf L}$ are  accessibile \cite{KurniawanThesis}. Analogously to the case of closed systems, the key ingredient of this result can be identified with the noncommutativity of the elements of   ${\mathbf L}$. But what about models 
where $\mathbf{L}$  includes only mutually commutating  Lindbladians? Is there a way to 
 expand their algebra 
$\mathfrak L({\mathbf L})$  to cover the full GKLS algebra? 
 For close quantum systems it has been observed that one can 
 substantially change the dimension of the dynamical Lie algebra $\mathfrak L({\mathbf H})$ through frequently observing 
a part of the system \cite{DanielZeno},  or by tampering it with a strong dissipative process that exhibits decoherence-free subspaces \cite{FullcontrolNoise}
 (the gain being exponential in some cases). 
As a matter of fact, on the basis of 
the 
quantum
Zeno effect \cite{PascazioFacchi1}, starting with a set of commuting control Hamiltonians ${\mathbf H}$,
noncommutativity can be enforced through 
frequently projecting out part of the system onto a subspace where accessibility and hence full 
controllability 
is
achieved. 
Also  it has been observed that the projection trick can be reversed: specifically, starting from a set of noncommuutative Hamiltonians ${\mathbf H}$
one can construct a new set $\tilde{\mathbf H}$ formed by commutative elements on an extended Hilbert space which under projection reduces to the original one. 
This mechanism was studied in great detail in \cite{HamiltonianPurification}, 
where, borrowing from the notion of purification of mixed quantum states \cite{NielseChuang},  the term \emph{Hamiltonian purification} was introduced.

One may then ask whether a similar procedures
can be applied to  the algebra of a set ${\mathbf L}$ of Lindbladians, 
namely, if it is possible to enlarge $\mathfrak L({\mathbf L})$  by means of some projection mechanisms and, on the contrary, if \emph{Lindbladian purification} is always achievable.
In this article we address these issues  showing that indeed 
any set ${\mathbf L}$ of  
Lindbladians
can be  ``purified,'' i.e., can be made commutative with each other, by embedding them in a larger space (note that the term ``pure'' was already used in \cite{Lindblad} for Markovian generators 
 in a slightly different way).
To this end, we need to employ a different scheme  from those for the Hamiltonian purification introduced in \cite{HamiltonianPurification}, since the naive application of the latter  trivially violates some structural properties of GKLS generators (more details in the following). 
Our construction allows 
one to make Lindbladians
and Hamiltonians commutative on an extended space by means of  an auxiliary system, which,  through frequent non-selective  measurements,
yields the original noncommutative dynamics.
Moreover, we present a universal pair of  
Lindbladians that generate
the full GKLS dynamical Lie algebra for generic system sizes, the analysis providing us with 
  a short and elementary proof of the generic accessibility \cite{KurniawanThesis}.
Applying hence the Lindbladian purification procedure  to such a universal set 
we then show that almost all open systems become accessible,
even though their generators are commutative with each other,
by performing frequent non-selective measurements on a part of the system. }

This article 
is organized as follows. 
Along the lines of \cite{HamiltonianPurification} we begin
in Sec.\ \ref{sec:LindbladianPurification} by reviewing the definition of Hamiltonian purification and presenting 
an explicit construction for purifying an arbitrary number of Lindbladians and Hamiltonians. 
In Sec.\ \ref{sec:accessibility} we consider the accessibility of controlled 
master equations. 
Concluding remarks are given in Sec.\ \ref{sec:Conclusions}, and some details on the derivation of the projected dynamics and the proof of accessibility are provided in the Appendices.

\section{Lindbladian purification and non-selective Zeno measurements}
\label{sec:LindbladianPurification}
To begin with we  
first
review the definition of Hamiltonian purification \cite{HamiltonianPurification}. 
Suppose that we have $n$ control Hamiltonians, which are switched on and off to steer a $d$-dimensional quantum system $Q$.
Let ${\mathbf H}=\{H_{1},\ldots,H_{n}\}$ be the set of 
the control
Hamiltonians acting on the Hilbert space $\mathcal H_{d}$ of $Q$,  and $\tilde{{\mathbf H}}=\{\tilde{H}_{1},\ldots,\tilde{H}_{n}\}$ be a corresponding set of Hamiltonians acting on an extended Hilbert space $\mathcal H_{d_{E}}$
of dimension $d_E\,(>d)$, 
which includes $\mathcal H_{d}$ as a proper subspace. 
We call $\tilde{\mathbf H}$ a purifying set of ${\mathbf H}$ if all the elements of $\tilde{\mathbf H}$ commute with each other, 
   \begin{equation}
   \tilde{H}_{i}\tilde{H}_{j} =  \tilde{H}_{j}\tilde{H}_{i},\quad\forall\,i,j\in\{1,\ldots,n\},
   \end{equation}
and they are related to those from ${\mathbf H}$ through 
\begin{equation}
\label{eq:projectedHamiltonians}
H_{j}=P\tilde{H}_{j}P,\quad\forall\,j\in\{1,\ldots,n\},	
\end{equation}
with $P$ being the  
projection onto $\mathcal H_{d}$. 
For a generic set ${\mathbf H}$ 
consisting
of $n$ linearly independent Hamiltonians it can be shown \cite{HamiltonianPurification} that there 
always exists an  $\tilde{\mathbf H}$ where the minimal dimension $d_{E}^{(\mathrm{min})}$ of the extended Hilbert space is bounded above by $d_{E}^{(\mathrm{min})}\leq nd$.
For instance for  the case  with  $n=2$ Hamiltonians $H_{1}$ and $H_{2}$, Proposition 1 of Ref.\ \cite{HamiltonianPurification} states that a purifying set can be constructed on $\mathcal H_{d_{E}}=\mathcal H_{d}\otimes \mathcal H_{d_{A}}$ with an auxiliary single qubit Hilbert space $\mathcal H_{d_{A}}$, 
the purifications and the projector  being 
\begin{eqnarray}
\tilde{H}_{1}&=&H_{1}\otimes \openone_{2}+H_{2}\otimes\sigma_{x},\nonumber    \\
\tilde{H}_{2}&=&H_{2}\otimes \openone_{2}+H_{1}\otimes\sigma_{x}, \label{eq:purificationHamDan}
\end{eqnarray}
\begin{equation}
P=\openone_{d}\otimes \frac{\openone_{2}+\sigma_{z}}{2},
\end{equation}
with $\sigma_x$, $\sigma_z$, and $\openone_{2}$ the Pauli and the identity operators of the auxiliary qubit, respectively.
The mapping $\tilde{H}_{j}\rightarrow H_j$  can finally be realized through the quantum Zeno effect \cite{PascazioFacchi1,ref:QZS} by frequently monitoring the extended system 
via a von Neumann measurement which projects the system onto $\mathcal H_{d}$, i.e.,
\begin{equation}
\label{ZenoLimitH}
\lim_{N\to\infty}(Pe^{-i\tilde{H}_{j}t/N}P)^{N}=e^{-iH_{j}t}P.
\end{equation}

The question arises if an analogous construction can be extended to the case of 
Lindbladians. 
Specifically  consider a set
${\mathbf L}=\{\mathcal L_{1},\ldots,\mathcal L_{n}\}$ 
of $n$  GKLS
generators operating on a target system $Q$,
\begin{equation}
\mathcal{L}_j=\mathcal{K}_j+\mathcal{D}_j,\quad
j\in\{1,\ldots,n\}, 
\label{eqn:Lindbladian}
\end{equation}
with $\mathcal K_j$ and $\mathcal D_j$ being the  Hamiltonian and dissipator contributions, i.e., the super-operators 
\begin{eqnarray}
\mathcal K_j({\cdots}) 
&=&{-i}[H_j,{}\cdots{}]
\label{eqn:unitarygen} , \\
\label{eq:dissipativepart}
\mathcal D_j({\cdots}) &=&
\sum_\alpha[
2L_{j,\alpha}({\cdots})  L_{j,\alpha}^{\dagger}
\nonumber \\[-3truemm]
&&\qquad
{}-L_{j,\alpha}^{\dagger}L_{j,\alpha}({\cdots})
- ({\cdots})  L_{j,\alpha}^{\dagger}L_{j,\alpha}
],
\end{eqnarray} 
 $L_{j,\alpha}$ being the Lindblad operators acting on the Hilbert space $\mathcal H_{d}$ of $Q$. 
We ask whether if it is possible  to associate with  $\mathbf{L}$ a purifying set $\tilde{\mathbf{L}}=\{\tilde{\mathcal L}_{1},\ldots,\tilde{\mathcal L}_{n}\}$ formed by GKLS generators
possibly acting on an extended system, which are mutually commuting, i.e.,
\begin{equation} \label{COMMPROP}
\tilde{\mathcal L}_i \circ \tilde{\mathcal L}_j = \tilde{\mathcal L}_j \circ \tilde{\mathcal L}_i,\quad\forall\,i,j\in\{1,\ldots,n\},
\end{equation} 
from which one can recover the original elements via a projective mapping that should mimics~(\ref{ZenoLimitH})
(in the above expressions we used the symbol ``$\circ$" to indicate the composition of super-operators). 

A natural guess for identifying  $\tilde{\mathbf{L}}$ and the projective  mapping would be to simply transporting the  purification schemes of Ref.\ \cite{HamiltonianPurification} at super-operator level, or equivalently, to represent the  ${\mathcal L}_j$s as operators in Liouville space \cite{ROY} 
 and then simply applying to them 
the Hamiltonian purication scheme. This simple trick however does not work because, for instance, mapping as (\ref{eq:purificationHamDan}) will take
positive operators into non-positive one, hence spoiling one fundamental property of GKLS generators. 
Another problem comes from the fact that  for  
Markovian open systems described by 
Lindbladians,
the quantum Zeno effect, which as we have seen  is responsible for the implementation of the mapping $\tilde{H}_{j}\rightarrow H_j$,
does not take place: a Markovian system can leak from one subspace specified by the projection operator belonging to a measurement outcome even in the limit of infinitely frequent projective measurements.
In spite of these issues however 
a Lindbladian purification scheme can   be obtained with the following simple construction:

\bigskip
\noindent
\textit{A purifying set $\tilde{\mathbf{L}}$ can always be constructed 
by introducing 
an auxiliary Hilbert space $\mathcal H_{n}$ of dimension $n$ and identifying 
the Hamiltonians $\{\tilde{H}_j\}$ and
the Lindblad operators $\{\tilde{L}_{j,\alpha}\}$ of the purifying element  $\tilde{\mathcal {L}}_j = \tilde{\mathcal {K}}_j  + \tilde{\mathcal {D}}_j$ as
\begin{equation}
\label{eq:purificationschemeMe}
\begin{cases}
\medskip
\tilde{H}_j=n H_j\otimes \ket{j}\bra{j},\\
\tilde{L}_{j,\alpha}=\sqrt{n} \, L_{j,\alpha}\otimes \ket{j}\bra{j},
\end{cases}
\quad j\in\{1,\ldots,n\},	
\end{equation}
with $\{\ket{j}\}_{j=1}^{n}$ being an orthonormal basis for $\mathcal H_{n}$.}

\bigskip
\noindent
Obviously through such a construction the operators $\{\tilde{H}_j\}$ and  $\{\tilde{L}_{j,\alpha}\}$ commute with each other for different $j$ trivially ensuring the
requirement~(\ref{COMMPROP}). 
Regarding the analog of (\ref{ZenoLimitH}) 
we focus on 
non-selective projective measurement \cite{Schwinger,SchwingerB} operating on the auxiliary system, i.e., the 
completely positive and trace preserving  (CPTP) mapping of the form
\begin{equation}
\label{eq:non-selectiveM}
\mathcal P({\cdots}) =\sum_{k}P_{k}({\cdots})  P_{k},
\end{equation}
given 
in terms of 
a 
complete 
set 
of 
orthonormal projection operators  
$\{P_{k}\}$ corresponding to measurement outcomes and satisfying
$P_kP_{k'}=\delta_{kk'}P_k$ and $\sum_{k}P_{k}=\openone$. 
Notice that if we perform $(N+1)$ of such non-selective measurements at regular time intervals $t/N$ during the evolution driven by a  
Lindbladian
$\mathcal{L}$, the system will evolve according to the CPTP transformation 
\begin{eqnarray}
\Phi_{t,N}^{(\cal L )} &:=& (\mathcal{P}\circ e^{\mathcal{L}t/N}\circ\mathcal{P})^N  \nonumber \\
&=&\left[\mathrm{id}+(\mathcal P\circ\mathcal L\circ \mathcal P)\tfrac{t}{N}+O\!\left(\tfrac{t^2}{N^{2}}\right)\right]^N\circ\mathcal{P},
\end{eqnarray} 
which in the limit of $N\rightarrow \infty$ converges to 
\begin{equation} 
{\Phi}_{t,\infty}^{(\cal L )}  =  \lim_{{N\to\infty}}\Phi_{t,N}^{(\cal L)} = 
e^{(\mathcal P\circ\mathcal L\circ \mathcal P)t}\circ \mathcal P,
\label{eq:ZenoLimitSemigroups}
\end{equation}
where $\mathrm{id}$ is the identity map and where we used the idempotent property $\mathcal P=\mathcal P\circ\mathcal P=\mathcal P^2$ of~(\ref{eq:non-selectiveM}).
Equation (\ref{eq:ZenoLimitSemigroups}) can also  be derived following a pertubative approach with a strong amplitude-damping channel inducing the projection $\mathcal P$ \cite{Zanardi1, Vittorio, Zanardi2}. 
In our construction Eq.~(\ref{eq:ZenoLimitSemigroups}) is the formal counterpart of the Zeno limit~(\ref{ZenoLimitH}): it shows that alternating the dynamics induced by a GKLS
generator $\mathcal{L}$ with ${\mathcal P}$ induces on the system an evolution which can be effectively described in terms of an effective dynamical generator described by the projected super-operator $\mathcal{P}\circ\mathcal{L}\circ\mathcal{P}$. It should be stressed that this last is not in GKLS form, i.e., it is not a Lindbladian. Indeed it acts as a proper
Lindbladian only within the subspace specified by the super-projector $\mathcal{P}$, but the map  $e^{(\mathcal{P}\circ\mathcal{L}\circ\mathcal{P})t}$ itself is not CPTP (an explicit example of this fact is provided in Appendix~\ref{APPANEW}). Still we are going to identify \eqref{eq:ZenoLimitSemigroups} with the mechanism that 
yields  the original  Lindbladians $\mathcal L_j \in \mathbf{L}$  expressed in the form (\ref{eqn:Lindbladian})--(\ref{eq:dissipativepart})  from their 
 purified counterparts $\tilde{\mathcal L}_j$ of with (\ref{eq:purificationschemeMe}). 
For this purpose we assume the projectors $P_k$ in (\ref{eq:non-selectiveM}) to be of the form 
\begin{equation} \label{DEFPK}
P_{k}=\openone\otimes \ket{\phi_k}\bra{\phi_k},
\end{equation}
where $\{\ket{\phi_{k}}\}_{k=1}^{d_{A}}$ is an orthonormal basis for the auxiliary Hilbert space which is chosen to be mutually unbiased \cite{MUTU} against the orthonormal basis $\{\ket{j}\}_{j=1}^{d_A}$ used for the purification (\ref{eq:purificationschemeMe}).
Then as shown in Appendix \ref{sec:derivationLP}  one can verify
that 
under the transformation~(\ref{eq:ZenoLimitSemigroups}) a generic density operator $\rho_Q(0)$  for the original system, obtained by taking the trace over the auxiliary Hilbert space $\mathcal{H}_{d_A}$, evolves according to 
\begin{equation}
\label{eq:reducedpurif}
\rho_{Q}(t)=e^{\mathcal L_jt }\rho_{Q}(0), 
\end{equation}
recovering hence  the original dynamics generated by the unpurified Lindbladian $\mathcal L_j$.

\section{Accessibility}
\label{sec:accessibility}
We now turn our attention to the question on how frequent non-selective measurements can enrich the algebra $\mathfrak L({\mathbf L})$ 
of a Markovian open quantum system described by a collection  $\mathbf L$ of controlled generators. 
Specifically we shall focus on systems driven by  master equations of the form 
\begin{equation}
\frac{\partial}{\partial t} {\rho}(t)=\mathcal{L}(t)\rho(t),
\label{eq:mastereq}
\end{equation}
where the super-operator  $\mathcal{L}(t)=\mathcal{K}(t)+\mathcal D$
is provided by a constant dissipative part represented by Lindblad operators $L_\alpha$,
and by a time-dependent Hamiltonian  term $\mathcal{K}(t)({\cdots}) =-i[H(t),{}\cdots{}]$ with 
 \begin{equation}
H(t)=H_{0}+\sum_{k=1}^{m}u_{k}(t)H_{k},
\end{equation}
$\{u_k(t)\}_{k=1}^m$ being  classical control fields  that can be operated to switch on and off $m$ control Hamiltonians $\{H_k\}_{k=1}^m$.
This corresponds to having a  control set  $\mathbf L:=\{\mathcal L_0,\mathcal K_{1},\ldots,\mathcal K_{m}\}$ consisting of a drift (unmodulated) term 
\begin{equation}
\mathcal{L}_0=\mathcal{K}_0+\mathcal{D},
\end{equation}
that includes both the dissipative part $\mathcal{D}$ and  the 
 Hamiltonian contribution $\mathcal{K}_0({\cdots}) =-i[H_0,{}\cdots{}]$, and of  the set of Hamiltonian control generators 
\begin{equation}
\mathcal{K}_k({\cdots}) =-i[H_k,{}\cdots{}]
,\quad k\in\{1,\ldots,m\}.
\end{equation}
As already mentioned in the introduction, for a closed quantum system, i.e., without the dissipative part $\mathcal D$,  the algebra $\mathfrak L({\mathbf L})$ associated
with  ${\mathbf L}$ (i.e., the set of  all real linear combinations and iterated commutators of these elements, drift term included) will 
fully characterize the set of unitary operations that can be implemented through shaping the control functions $\{u_k(t)\}_{k=1}^m$.
For an open quantum system described by the master equation \eqref{eq:mastereq}, instead, $\mathfrak L({\mathbf L})$ 
only characterizes the accessibility of the system. 
For a detailed analysis of the general  structure of  $\mathfrak L({\mathbf L})$  and simple examples, we refer to \cite{KurniawanThesis}. 
Here we focus instead on studying how the purification mechanism can influence the dimension of $\mathfrak L({\mathbf L})$. In particular we shall 
see how a set of commutative Lindbladians can be  turned into a new set of noncommuutative Lindbladians which  grant accessibility to the full GKLS algebra via the projection through frequent measurements.

To show this we start by showing that it is possible to identify  a set ${\mathbf L}$ formed by just a pair of Lindbladians whose algebra  $\mathfrak L({\mathbf L})$ spans the full GKLS algebra. 
We therefore first prove that the pair
\begin{gather}
\label{eq:gen1}
\mathcal L_0=-i\ad_{H_{0}}{}+\mathcal D_{\ket{1}\bra{2}},\\
\label{eq:gen2}
\mathcal K=-i\ad_{\ket{1}\bra{1}}
\end{gather}
with 
\begin{equation}
H_{0}=\sum_{j=1}^{d-1}\ket{j}\bra{j+1}+\mathrm{h.c.},
\end{equation}
where 
\begin{gather}
-i\ad_H({\cdots}) =-i[H,{}\cdots{}],
\label{eqn:NotationI}
\\
\mathcal D_L({\cdots}) 
=2L({\cdots})  L^\dag
-L^\dag L({\cdots}) 
-({\cdots}) L^\dag L,
\label{eqn:NotationII}
\end{gather}
does the job, namely, 
every possible Lindbladian can be generated by linear combinations and iterated commutators of \eqref{eq:gen1} and \eqref{eq:gen2}. 
We only sketch the main steps here, whereas the details can be found in Appendix \ref{sec:accesibleLindP}\@. 
In the following we also use the notations
\begin{gather}
\mathcal{D}_{A,B}({\cdots}) =2B({\cdots})  A^\dag-A^\dag B({\cdots}) -({\cdots})  A^\dag B,
\label{eqn:NotationIII}
\\
\Ad_U({\cdots}) =U({\cdots}) U^\dag.
\label{eqn:NotationIV}
\end{gather}

We first note that terms of the form $-i\ad_{\ket{j}\bra{j}}$ commute with the dissipative part $\mathcal D_{\ket{1}\bra{2}}$ and according to \cite{SIdentification} we can generate every element in $-i\ad_{\mathfrak{u}(d)}$ with $\mathfrak{u}(d)$ the Lie algebra of $d\times d$ hermitian matrices. 
Using $\Ad_{U(d)}=\exp(-i\ad_{\mathfrak{u}(d)})$ with $U(d)$ being the unitary group, we can get 
\begin{equation}
\Ad_{U}{}\circ\mathcal D_{\ket{1}\bra{2}}\circ\Ad_{U^{\dagger}}=\mathcal D_{U\ket{1}\bra{2}U^{\dagger}}
\end{equation}
for any $U\in U(d)$.  
Now we consider unitaries $U$ that act as $U\ket{j}=\sum_{k\in\mathcal I}c_{k}^{(j)}\ket{k}$ for $j=1,2$ and $\mathcal I=\{1,2,3,4\}$. 
We numerically verified that, from $\mathcal D_{U\ket{1}\bra{2}U^{\dagger}}$, thus created together with $-i\ad_H$ for all Hamiltonians $H=\sum_{i,j\in\mathcal I}h_{ij}\ket{i}\bra{j}$ having support on $\mathcal I$, all the operators of the form 
\begin{equation}
\begin{cases}
\medskip
\hphantom{i(}
\mathcal D_{\ket{i}\bra{j},\ket{k}\bra{l}}+\mathcal D_{\ket{k}\bra{l},\ket{i}\bra{j}},\\
i(\mathcal D_{\ket{i}\bra{j},\ket{k}\bra{l}}-\mathcal D_{\ket{k}\bra{l},\ket{i}\bra{j}}),
\end{cases}
\quad
i,j,k,l\in\mathcal I
\label{eq:basisop2}
\end{equation}
can be generated. 
Doing the same for different quartets $\mathcal I=\{i,j,k,l\}$ we are able to provide linearly independent operators \eqref{eq:basisop2} for all $i,j,k,l\in\{1,\ldots,d\}$. 
Since any Lindbladian can be written in the Kossakowski form as a linear combination of those operators, it means that every Lindbladian can be generated through iterated commutators and linear combinations of the pair of generators  $\{\mathcal{L}_0,\mathcal{K}\}$ in \eqref{eq:gen1} and \eqref{eq:gen2}. Given that this specific pair of Lindbladians is accessible, it then follows from the standard argument (see, e.g., \cite{DanielZeno}) that almost all pairs are. 
This was shown previously in a more abstract way by Kurniawan \cite{KurniawanThesis}.

Now that we have found a pair $\mathbf L= \{\mathcal{L}_0,\mathcal{K}\}$ that describes an accessible quantum system in arbitrary dimensions, we can make them commutative using a two-dimensional ($d_{A}=2$) auxiliary  Hilbert space, i.e., we can purify them to
\begin{gather}
\tilde{\mathcal L}_0=-2i\ad_{H_{0}\otimes\ket{2}\bra{2}}{}+\sqrt{2}\, \mathcal D_{\ket{1}\bra{2}\otimes\ket{2}\bra{2}},\\
\tilde{\mathcal K}=-2i\ad_{\ket{1}\bra{1}\otimes\ket{1}\bra{1}}.
\end{gather}
Obviously on the extended Hilbert space the Lie algebra associated with the set $\tilde{\mathbf L}= \{\tilde{\mathcal{L}}_0,\tilde{\mathcal{K}}\}$ is just two-dimensional, $\dim\mathfrak{L}(\tilde{\mathbf L})=2$, and the system is not accessible. 
If we perform frequent non-selective projective measurements on the auxiliary system described by the superprojector \eqref{eq:non-selectiveM} with $P_\pm=\openone \otimes \ket{\pm}\bra{\pm}$, where $\ket{\pm}$ are defined at the end of Sec.\ \ref{sec:LindbladianPurification}, the original dynamics is recovered as \eqref{eq:reducedpurif} and the system becomes accessible. 
The existence of such a specific setup allows us to conclude \cite{DanielZeno} that \emph{almost all open quantum systems become accessible by Zeno measurements}.

\section{Conclusions}
\label{sec:Conclusions}
We have generalized the work \cite{HamiltonianPurification} on Hamiltonian purification by establishing a new and simple purification scheme for Lindbladians, which is also applicable to Hamiltonians. 
Given $n$ Lindbladians, they can be made commutative by adding an $n$-dimensional auxiliary system to extend the Lindblad operators with hermitian projectors that form an orthonormal basis for the auxiliary space. 
Through the projection by Zeno measurements for semigroup dynamics the original possibly noncommutative dynamics can be recovered by frequently measuring the auxiliary system in a non-selective way. 
Moreover, we have proven that the pair of Lindbladians \eqref{eq:gen1} and \eqref{eq:gen2} describes an accessible open quantum system for generic system sizes, which tells us that generally a nonaccessible open quantum system is turned into an accessible one by frequent non-selective measurements.
The model has also potential applications in simulating an arbitrary Markovian open system dynamics \cite{OpenSySim, OpenSySim2} by steering it through control fields.

Clearly, the presented purification scheme also works for observables and density operators, although, except for the partial trace, an operational way that allows us to recover the original observables and states is not known to us. 
Since the noncommutativity is a unique feature of quantum mechanics, and in fact it was argued in \cite{Herschclassical1, Herschclassical2} that the noncommutativity distinguishes between quantum and classical mechanics, it is tempting to say that every quantum system can be made classical by purifying it to a larger space.

\acknowledgments
We like to thank John Gough for fruitful discussions. 
This work was supported by the Top Global University Project from the Ministry of Education, Culture, Sports, Science and Technology (MEXT), Japan.
DB acknowledges support from EPSRC grant EP/M01634X/1\@.
KY was supported by the Grant-in-Aid for Scientific Research (C) (No.\ 26400406) from the Japan Society for the Promotion of Science (JSPS) and by the Waseda University Grant for Special Research Projects (No.\ 2016K-215). HN was supported by the Waseda University Grant for Special Research Project (No.\ 2016B-173).

\appendix
\section{Projected Lindbladians}
\label{APPANEW} 
As an  example of the fact that the projected counterpart $\mathcal{P}\circ\mathcal{L}\circ\mathcal{P}$ of a Lindbladian 
$\mathcal L$ does not generate proper quantum dynamics, consider for instance the case where  ${\mathcal L}$ describes a qubit amplitude damping with fixed point $|0\rangle\langle 0|$ (this is characterized by a null Hamiltonian term $H=0$ and a unique Lindblad operator $L_\alpha = |0\rangle\langle 1|$) and 
where the transformation $\mathcal{P}$ is the dephasing map \cite{NielseChuang} associated with the canonical qubit base,  i.e.,
\begin{equation}
\mathcal{P}({\cdots}) 
=\ket{0}\bra{0}({\cdots})\ket{0}\bra{0}
+\ket{1}\bra{1}({\cdots})\ket{1}\bra{1}.
\end{equation}
Accordingly for an arbitrary density matrix $\rho$ we have 
\begin{equation}
\lim_{t\rightarrow\infty}(e^{(\mathcal{P}\circ\mathcal{L}\circ\mathcal{P})t}\circ\mathcal{P})\rho
=\ket{0}\bra{0}, 
\end{equation}
while on the contrary 
\begin{eqnarray}
\lim_{t\to\infty}
e^{(\mathcal{P}\circ\mathcal{L}\circ\mathcal{P})t}\rho
&=&\ket{0}\bra{0}
+(\mathrm{id} -\mathcal{P})\rho
\nonumber\\
&=&\ket{0}\bra{0}
+\ket{0}\bra{0}\rho\ket{1}\bra{1}
+\ket{1}\bra{1}\rho\ket{0}\bra{0},
\nonumber\\
\end{eqnarray}
which in general is not a valid state.

\section{Derivation of the Projected Dynamics}
\label{sec:derivationLP}
We start by noticing that given a generic  non-selective transformation $\mathcal P$ as in (\ref{eq:non-selectiveM}) and the  unitary generator  
$\mathcal{K}$ with Hamiltonian $H$, 
the following  identity holds
\begin{equation} \label{HAMP} 
(\mathcal{P}\circ\mathcal{K}\circ\mathcal{P})({\cdots}) =-i\sum_k[H^{(k)},P_k({\cdots}) P_k], 
\end{equation}
where $H^{(k)}=P_k HP_k$. Similarly, given a dissipator $\mathcal{D}$  characterized by Lindblad operators $L_{\alpha}$  we have 
\begin{eqnarray}
&&
(\mathcal P\circ\mathcal{D}\circ\mathcal P)({\cdots}) 
\nonumber\\
&&\qquad
= 
\sum_{\alpha, k,k'}\Bigl[
2L_\alpha^{(kk')}({\cdots}) L_\alpha^{(kk')\dag} 
- L_\alpha^{(kk')\dag}   L_\alpha^{(kk')}  ({\cdots})  P_{k'}
\nonumber\\[-3truemm]
&&\qquad\qquad\qquad\qquad\qquad\qquad\quad\ \,%
{}-P_{k'}({\cdots}) L_\alpha^{(kk')\dag}   L_\alpha^{(kk')} 
\Bigr],
\nonumber\\
\label{eq:projectedDissipator}
\end{eqnarray}
with $L_\alpha^{(kk')}  = P_kL_\alpha P_{k'}$. 
Assume next $H$ and $L_\alpha$ as those associated with the Lindbladian $\tilde{\mathcal {L}}_j$ with (\ref{eq:purificationschemeMe}), and 
$P_k$ as in (\ref{DEFPK}). Since $\{|\phi_k\rangle\}$ is mutually unbiased with respect to $\{ |j\rangle\}$ the following identity holds,
\begin{equation} 
\langle j | \phi_k\rangle = e^{- i\varphi_{jk}}/\sqrt{d_A}, 
\end{equation} 
with $\varphi_{jk}$ generic phases,  and hence 
\begin{eqnarray}
&&\tilde{H}_j^{(k)}= P_k \tilde{H}_j P_k= H_j \otimes \ket{\phi_k}\bra{\phi_k} = H_j P_k,  \\
&&\tilde{L}_{j,\alpha}^{(kk')} =  P_k\tilde{L}_{j,\alpha} P_{k'} =e^{i (\varphi_{jk}-\varphi_{jk'})}  L_{j,\alpha}  \otimes \ket{\phi_k}\bra{\phi_{k'}}/\sqrt{d_A},
\nonumber\\
\\
&&\tilde{L}_{j,\alpha}^{(kk')\dag} \tilde{L}_{j,\alpha}^{(kk')}  =L_{j,\alpha}^\dag L_{j,\alpha} \otimes \ket{\phi_{k'}}\bra{\phi_{k'}}=  L_{j,\alpha}^\dag L_{j,\alpha}  P_{k'}/d_A.  
\nonumber\\
\end{eqnarray}
Inserting these into~(\ref{HAMP}) and (\ref{eq:projectedDissipator}) we then  obtain  
\begin{eqnarray} \label{FIN1}
(\mathcal{P}\circ\tilde{\mathcal{K}}_j\circ\mathcal{P})({\cdots}) &=&
 (\mathcal{K}_j \circ \mathcal{P})({\cdots}),  \\
( \mathcal{P}\circ\tilde{\mathcal{D}}_j \circ\mathcal{P})({\cdots}) &=& (\mathcal{D}_j \circ \mathcal{P})({\cdots}) 
\nonumber\\
&&{}+2  \sum_{\alpha} 
L_{j,\alpha} \mathcal{T}({\cdots})  L_{j,\alpha}^\dag,  
\end{eqnarray}
that is 
\begin{equation} 
( \mathcal{P}\circ\tilde{\mathcal{L}}_j \circ\mathcal{P})({\cdots}) 
 = (\mathcal{L}_j \circ \mathcal{P})({\cdots}) +2  \sum_{\alpha} 
L_{j,\alpha} \mathcal{T}({\cdots})  L_{j,\alpha}^\dag, \label{IMPO} 
\end{equation}
where  $\mathcal{T}$ is the super-operator 
\begin{equation} 
\mathcal{T}({\cdots})= \frac{1}{d_A}\openone_A\Tr_A[({\cdots})]- \mathcal{P}({\cdots}),\end{equation} 
with  $\Tr_A[({\cdots})]$ indicating the partial trace over the auxiliary system $A$ and $\openone_A$ being the identity operator on the associated Hilbert space.

In order to prove~(\ref{eq:reducedpurif}) let us now focus on the evolution induced by CPTP map ${\Phi}_{t,\infty}^{(\mathcal{L}_j)}$ in (\ref{eq:ZenoLimitSemigroups})
associated with the $j$th element of  $\mathbf L$ on a generic 
 density matrix $\rho(0)$ of the joint system $Q+A$, i.e., $\rho(t) = {\Phi}_{t,\infty}^{(\mathcal{L}_j)}\rho(0)$. We are interested in  the dynamics of the reduced density matrix of 
 $Q$, i.e.,  
\begin{equation} 
\rho_Q(t) =  \Tr_A[\rho(t)] = \Tr_A[{\Phi}_{t,\infty}^{(\mathcal{L}_j)}\rho(0)] .
\end{equation} 
By taking the first derivative with respect to $t$ and using (\ref{IMPO}) we obtain 
\begin{eqnarray} 
\frac{\partial}{\partial t}  \rho_Q(t)& =&  \Tr_A[ (\mathcal{P}\circ\tilde{\mathcal{L}}_j \circ\mathcal{P})\rho(t)] \nonumber \\
&=& \mathcal{L}_j(\Tr_A[\mathcal{P}\rho(t)]) + 2 \sum_{\alpha} 
L_{j,\alpha}\Tr_A[ \mathcal{T}\rho(t)]   L_{j,\alpha}^\dag, \nonumber\\
\end{eqnarray} 
which finally yields the thesis 
\begin{equation} 
\frac{\partial}{\partial t}  \rho_Q(t) =  \mathcal{L}_j\rho_Q(t) \; \Longrightarrow\; \rho_Q(t)=e^{\mathcal{L}_j t} \rho_Q(0), 
\end{equation} 
by noticing that 
\begin{equation}
\Tr_A[ \mathcal{P}\rho(t)] = \rho_Q(t),\quad  
\Tr_A[ \mathcal{T}\rho(t)] = 0. 
\end{equation}

\section{An Accessible Pair of Lindbladians}
\label{sec:accesibleLindP}
Here we show that the pair of Lindbladians $\{\mathcal{L}_0,\mathcal{K}\}$ in \eqref{eq:gen1} and \eqref{eq:gen2} generates an accessible system.
We use the notations \eqref{eqn:NotationI}--\eqref{eqn:NotationIV}.
First of all, we show that $\mathcal{K}=-i\ad_{\ket{j}\bra{j}}$ commutes with the dissipative part $\mathcal{D}_{\ket{1}\bra{2}}$ of $\mathcal{L}_0$ in \eqref{eq:gen1}.
Using an identity \cite{Shai}
\begin{equation}
[-i\ad_{H},\mathcal D_A]=\frac{1}{2}\mathcal D_{A-i[H,A]}-\frac{1}{2}\mathcal D_{A+i[H,A]},
\end{equation} 
we have
\begin{eqnarray}
[-i\ad_{\ket{j}\bra{j}},\mathcal D_{\ket{1}\bra{2}}] 
&=&\frac{1}{2}\mathcal D_{\ket{1}\bra{2}-i\ket{1}\bra{2}\delta_{j1}+i\ket{1}\bra{2}\delta_{j2}} \nonumber \\
&&
{}-\frac{1}{2}\mathcal D_{\ket{1}\bra{2}+i\ket{1}\bra{2}\delta_{j1}-i\ket{1}\bra{2}\delta_{j2}}.	\nonumber\\
\end{eqnarray}
For $j\neq 1,2$ it trivially vanishes,
while for $j=1,2$ we get $\frac{1}{2}(|1\mp i|^{2}-|1\pm i|^{2})\mathcal D_{\ket{1}\bra{2}}=0$, where we have used $\mathcal D_{\alpha A}=|\alpha|^{2}\mathcal D_A$.
This commutativity implies that we can generate every $-i\ad_{\mathfrak u(d)}$ (see \cite{SIdentification}) and thus every $\Ad_{U}{}\circ\mathcal D_{\ket{1}\bra{2}}\circ\Ad_{U^{\dagger}}=\mathcal D_{U\ket{1}\bra{2}U^{\dagger}}$ for any $U\in U(d)$. 
Taking unitaries $U\ket{j}=\sum_{k\in\mathcal I}c_{k}^{(j)}\ket{k}$ for $j=1,2$ and $\mathcal{I}=\{1,2,3,4\}$, we have
\begin{equation}
\label{eq:lincomb}
\mathcal D_{U\ket{1}\bra{2}U^{\dagger}}
=\sum_{i,j,k,l\in \mathcal I}(c_{i}^{(1)})^{*}c_{j}^{(2)}c_{k}^{(1)}(c_{l}^{(2)})^{*}\mathcal D_{\ket{i}\bra{j},\ket{k}\bra{l}}.	
\end{equation} 
We numerically verified 
that all the operators of the form \eqref{eq:basisop2} for $\mathcal{I}=\{1,2,3,4\}$ can be obtained by linear combinations of $\mathcal{D}_{U\ket{1}\bra{2}U^\dag}$ and $-i\ad_H$ with different $U$ and $H=\sum_{i,j\in\mathcal I}h_{ij}\ket{i}\bra{j}$ on $\mathcal{I}=\{1,2,3,4\}$. 
The same argument applies to any quartets $\mathcal{I}=\{i,j,k,l\}$, and all the operators of the form \eqref{eq:basisop2} for all $\mathcal{I}$ are available.
Then, every Lindbladian can be given as a linear combination of those operators, i.e.,
\begin{eqnarray}
\mathcal L=\sum_{i,j,k,l}
[&&
c_{ijkl}^{+}(\mathcal D_{\ket{i}\bra{j},\ket{k}\bra{l}} + \mathcal D_{\ket{k}\bra{l},\ket{i}\bra{j}}) \nonumber \\[-2truemm]
&&{}+c_{ijkl}^{-}i(\mathcal D_{\ket{i}\bra{j},\ket{k}\bra{l}} - \mathcal D_{\ket{k}\bra{l},\ket{i}\bra{j}})]
\end{eqnarray}
with some coefficients $c_{ijkl}^\pm$.

\end{document}